\documentstyle[aps,preprint]{revtex}
\begin{document}
\title{Ground-State Roughness of the Disordered Substrate and Flux Lines
  in $d=2$ } 
\date{Submitted 28 May 1996 - to appear in PRL}
\author{Chen Zeng$^{a}$, A. Alan Middleton$^{a}$ and Y. Shapir$^b$}
\address{$^a$ Department of Physics, Syracuse University,
Syracuse, NY 13210, USA} 
\address{$^b$ Department of Physics and Astronomy, University of Rochester,
Rochester, NY 14627, USA}

\maketitle 

\begin{abstract}

We apply optimization algorithms to the problem of finding
ground states for crystalline surfaces and flux lines
arrays in presence of disorder.
The algorithms provide ground states
in polynomial time, which provides for a more precise study of
the interface widths than from Monte Carlo simulations at finite
temperature.  Using $d=2$ systems up to size $420^2$, with a minimum
of $2 \times 10^3$
realizations at each size, we find very strong evidence for
a $\ln^2(L)$ super-rough state at low temperatures.
\end{abstract}
%
%


The flux-lines array formed in a 2D type-II dirty superconductor
with the magnetic field parallel to the plane and the surface
configurations of a crystalline defect-free material
deposited on a disordered substrate (DS)
are closely related systems.
They have been studied both for the intrinsic interest and because
they serve as prototypical models for elastic media in a disordered
environment.
Both have low temperature glassy phases in which equilibrium
and dynamic properties are dominated by the disorder. In the
continuum limit they are both described by the random-phase sine-Gordon
model (RSGM).
However, analytic
attempts at understanding the equilibrium properties of the
glassy phase based on the RSGM have yielded conflicting results
\cite{RGcalc,variationalcalc}.
Finite temperature simulations of the RSGM or the corresponding
discrete Gaussian model for the disordered substrate have also been
ambiguous \cite{BatrouniHwa,Rieger,Lancaster,CuleShapir,MarinariEtal}.
Moreover, it is not clear to which extent universality
arguments, which yield the RSGM as the continuum limit,
can be trusted at temperatures well below the glass transition.

The aim of the present work is to address both issues by finding
the exact minimum energy configurations in discrete models
of the DS surface and that of the flux-lines array. This yields 
their T=0 shapes for any given disorder realization; averaging
over disorder allows for
the evaluation of their averaged physical properties.
In particular, the quantities that theory and simulation has focused on
are the height-height correlations in the disordered substrate model
and the line-line correlations in the flux-line model.
The flux-lines system is
discretized by its mapping to a surface model (see below) and one may
observe the transition by inspecting the roughness scaling of the respective
surfaces instead.
The transition into the glassy phase is exhibited by a change in these
height-height correlations in both models.

The predictions from the analytic studies of the RSGM are as follows:
Above the transition temperature $T=T_g$ the surface is always
logarithmically rough,
with a prefactor proportional to T.
Below the transition temperature, renormalization group (RG)
calculations \cite{RGcalc} predict
a $\ln^2(L)$ behavior, while variational approaches \cite{variationalcalc}
predict the persistence of the
$\ln(L)$ behavior but with the prefactor {\em unchanging}
for $T \le T_g$.

Numerical simulations have differed in their results as well. Simulations
of the RSGM with weak coupling have shown
no transition at all \cite{BatrouniHwa}.
Others have shown a transition with a $\ln(L)$ \cite{Rieger}
or a $\ln^2(L)$ behavior \cite{Lancaster}.
MC simulations of the discrete Gaussian version of the model have
exhibited the transition but the behavior of the roughness could fit better
the $\ln(L)$ \cite{CuleShapir} or the $\ln^2(L)$ \cite{MarinariEtal}, 
depending on the way the data was interpreted.

As shown below we find very strong evidence that the ground states
configurations for both systems exhibit a $\ln^2(L)$ roughness. This
reinforces the RG predictions and, moreover, indicates that they
hold all the way to T=0, a much wider temperature range than might be
anticipated.  We are able to make strong conclusions since we
are able to calculate averages over many more realizations and at larger
length scales.  As we show below, studying the correlation function
in wavevector space also allows for a clearer distinction between the two
predictions.

To be specific, the ground-state properties of 
two interface models with quenched random impurities 
are investigated in this letter. The first model, i.e., the 
disordered-substrate surface model\cite{CuleShapir}, is described by 
height variables $h_i$ defined on every site $i$ of a   
square lattice. The coupling between height variables is governed 
by the following Hamiltonian:     
\begin{equation}
H_1 = {\kappa}\sum_{<i,j>} |h_i - h_j| ,  
\label{e1} 
\end{equation}
\noindent where $\kappa $ is the step energy, and the summation runs 
over all nearest-neighbor bonds of the square lattice.
(The discrete Gaussian model usually
replaces this Hamiltonian near and above $T_c$ where they
belong to the same universality class).
In the case of a flat substrate the $h_i$
takes integer values. To model the disordered substrate, however, 
the height $h_i$ takes the values $h_i=n_i+d_i$ where 
$n_i$ is an integer and the quenched random
height $d_i$ is chosen uniformly (and independently for each site)
in the interval $(-1/2,+1/2]$. Clearly, any surface configuration
can be viewed as an interface along (001)-direction of a simple
cubic lattice with the substrate being identified with the basal 
plane. 
 
The second model is the Triangular Ising Solid-on-Solid model 
(TISOS)\cite{SOS1} where the surface configuration is 
described by height variables $h_i$ which are defined on 
every site $i$ of a triangular lattice and take only integer 
values. The Hamiltonian is chosen to be of the form:
\begin{equation} 
H_2 = \sum_{<ij>} J_{ij} (|h_i-h_j|-1) ,    
\end{equation}  
where the summation again runs over all nearest-neighbor bonds, 
and the coupling constants $J_{ij}$ are chosen uniformly (and 
independently for each bond) in the interval $[0,1]$ 
to simulate the quenched point disorder. 
In contrast to the first model, further 
constraints are imposed on the height variables: 
(1) the height difference $|h_i-h_j|$ for every bond must be  
either $1$ or $2$, and (2) the total height increment (clockwise or 
counterclockwise) along any elementary triangle is zero. 
See Fig.\ 1 for an example of this construction. It is    
straightforward to check that the SOS surface so defined describes 
an interface along the (111)-direction of a simple cubic lattice. 

There exists yet another well-known representation of the TISOS model
\cite{SOS1}, 
i.e., the dimer-covering model on a hexagonal lattice, which not only 
makes transparent the equivalence of the TISOS to an array 
of self-avoiding fluxlines in $d=2$ dimensions but also facilitates 
the implementation of an efficient network optimization algorithm 
to study the ground-state properties of the TISOS model. Here each 
bond on the triangular lattice with $|h_i - h_j|= 2$ is identified 
with a dimer on the dual bond of the hexagonal lattice. See 
Fig.\ 1 for a diagram of this construction. A simple inspection  
shows that a surface configuration $\{h_i\}$ is mapped onto 
a complete dimer covering of the hexagonal lattice.
The fluxline representation results from the
transition graph (superposition of the
two dimer-covering configurations) between an arbitrary 
dimer-covering and a fixed reference dimer-covering where all vertical 
bonds of the hexagonal lattice are covered with dimers.
Here the vertical direction then specifies the
direction of the fluxlines (or 
the direction of an external magnetic field).  See Fig.\ 1 
for an explicit illustration.
Let $E_r$, $E_d$ and $E_f$ denote 
the sum of the energies of all vertical bonds, 
of all bonds that are covered by dimers, and of all bonds 
that coincide with the fluxlines, respectively.
The following equality then
holds: $E_d(\{J_{ij}\}) - E_r(\{J_{ij}\}) = E_f(\{K_{ij}\}) - n L $,
where $n$ signifies the number of fluxlines, $L$ is the length 
of the system (see Fig.\ 1), and $K_{ij}$ denotes
a transformed set of random bonds according to $K_{ij}=J_{ij}$ 
for all non-vertical bonds, and $K_{ij}=1-J_{ij}$ for all vertical bonds.
This simple exercise therefore demonstrates that all different 
representations (interface along (111)-direction, dimers, fluxlines) 
are also energetically equivalent with a suitable transformation 
of the random bonds that still maintains its uniform and independent 
distribution. 

For the purpose of comparison, we now briefly summarize the known 
results on the ground-state properties of the above two models 
in the absence of disorder. In the DS model, the surface is trivially
flat (constant $h_i$). In the latter case (TISOS), however, the ground      
state is highly degenerate with a macroscopic entropy per
lattice site\cite{pureentropy}. 
By exploring the equivalence between the ground-state ensemble 
of TISOS and that of the antiferromagnetic Ising model on a triangular
lattice (thus the name TISOS), Bl\"ote and Hilhorst\cite{SOS2} have shown that 
the ground state is logarithmically rough due to the entropic fluctuations.
The structure factor $S(k) = \langle |h({\vec k})|^2 \rangle$
in the long wavelength limit
is given by
$S(k) = (K_s{\vec k}^{2})^{-1} $,
with the stiffness constant $K_s=\pi/9$.
The width $W$ of the surface then satisfies
$W^2 \equiv \{ h^2({\vec r}) \}  - \{ h({\vec r}) \}^2 
= (\pi K_s)^{-1} \ln(L) + \cdot\cdot\cdot$ for $L \rightarrow \infty$,
where $L$ is the linear size of
the system.
Here we have used $\{\cdot\}$ for the average over lattice sites, 
and $\langle \cdot\rangle$ for the ground-state ensemble average.  

The problem of finding the ground states of the above models in the 
presence of disorder is that of minimizing the total energy given by 
Eq. (1) and Eq. (2) respectively.
Different representations of the models 
naturally lend themselves to various efficient network optimization 
algorithms, of which we employed the network maximum-flow algorithm  
and the minimum-cost perfect matching algorithm\cite{FP} by utilizing 
the interface and dimer covering representations respectively.  

In the case of the interface representation, we apply the algorithm
presented in \cite{Alan}.
For a directed graph with given ``edge capacities'' indicating 
the maximum amount of fluid that can flow from one node to another
along the directed edge, the maximum flow algorithm determines
the maximum amount of flow that can be sustained between two given nodes,   
the source and the sink, given flow conservation at all
other nodes. It is easily shown that the maximum flow value is 
identical to the value of the minimum ``cut'',
which is defined as the weight of the set of edges
with minimum total capacities 
that, when cut, disconnect the source and the sink\cite{FP}.
This comes about because the minimum cut is the ``bottleneck''
through which all flow must pass.
To apply this algorithm,
a finite 3-dimensional graph is constructed
whose minimum cut directly 
corresponds to the minimum energy interface in the height model.
For the DS and TISOS models, graphs can be constructed whose minimum
cut surfaces lack overhangs and minimize the appropriate energies.
For example, the graphs for the TISOS model have the same topology
as for the random-bond Ising model (RBIM) as described in \cite{Alan}, but
the disorder is periodic in the $(111)$ direction (in RBIM, the disorder
is uncorrelated).
Once the corresponding graph is constructed, the
max-flow/min-cut algorithm can be directly applied.
Specifically, the push-relabel 
algorithm and code developed by Cherkassky and Goldberg
\cite{ChGo}
was adopted to the sparse graph considered here. 

We also employed another network optimization algorithm, namely, the 
minimum-cost perfect matching problem\cite{FP}.
We preform a direct calculation based on the dimer representation.
As discussed above,
this matching problem is that of finding a set of dimers that cover the
lattice with minimum cost.
The ground-state problem 
of the TISOS model is precisely of this type when formulated in terms of  
dimer coverings since such
coverings are perfect matchings on the hexagonal lattice.
Moreover, unlike the 3-dimensional graphs needed in flow algorithm, the 
graph here remains planar and much larger systems can be studied
with the same computer resources;
this allows us to more
clearly distinguish $\ln^2(L)$ from $\ln(L)$ roughness in the TISOS
model than the min-cut algorithm.
Standard algorithms \cite{FP, LEDA} are sufficient to solve the
problem for smaller systems.
We have developed a heuristic algorithm, making use of 
routines from the C++ LEDA library\cite{LEDA} to implement the
minimum-cost dimer matching algorithm.
The final sequential code, running on a single 
IBM RS/6000-390 workstation, takes about $1$, $15$ and $25$ minutes 
to find the exact ground-state for system sizes of $L=192, 420$, and 
$540$ respectively for one instance of disorder realization.     
Details of the algorithm's implementation will be 
presented elsewhere\cite{perfect}. 

The interface widths, configurational energy, and structure
factor were calculated for the ground states for
samples of various sizes.  These quantities
were then averaged over at least $2000$ samples of each size.
The largest length scales for which these many samples
were studied were $L=200, 420$ for the DS and TISOS models,
respectively.
The sample-averaged
results of our numerical calculations are summarized in
Fig.\ 2 and Fig.\ 3.

Fig.\ 2 shows the averaged 
squared sample widths $\overline{W^2} = \overline{\{h^2\} - \{h\}^2}$
as a function of sample size.
The $\overline{W^2}(L)$ curve is not fit well by a straight
line on this logarithmic-linear scale: for both the DS
and TISOS models, such a fit gives a line well outside the statistical
error bars for any range of data over more than a factor of 2 in $L$.
This is evidence that the proposal that
$\overline{W^2}(L) \sim \ln(L)$ does not
correctly describe the ground state.
The data is however fit quite well by the form
\begin{equation}
\label{fitform}
\overline{W^2}(L) = C + A\ln(L) + B\ln^2(L),
\end{equation}
with the constants being given by
$C= 0.022, 0.39$,
$A= 0.023, 0.34$,
and $B= 0.0064, 0.060$,
for the DS and TISOS
models, respectively.
The fact that these coefficients satisfy $A = 2(BC)^{1/2}$
to within a $10\%$ error
indicates that the width $\overline{W}$ is well fit by the form
$\overline{W} = C^{1/2} + B^{1/2} \ln(L)$ even for $L \approx 10$.

We find that even stronger evidence for the super-rough phase with
$\overline{W^2}(L) \sim \ln^2(L)$ is found by examining the sample-averaged
structure factors
$\overline{S(k)}$.
These structure factors must of course
be consistent with the $\overline{W^2}$ data, but
the shape of these curves for different $L$
shows that the structure factor data have converged
well for the sample sizes we examine, so that the curvature of
the $\overline{W^2}$ plots on a logarithmic scale is not an artifact of small
system sizes.
Plots of $k^2\overline{S(k)}$ are shown in Fig.\ 3
for the DS and TISOS models.
The structure factors
are found to be isotropic for small $k$;
this is indicated by the convergence of $\overline{S(k)}$ to a single
value for all $\vec{k}$ with the same magnitude $k$ as $k\rightarrow 0$.
If the coefficient $B$ in Eq.\ (\ref{fitform}) were zero,
$k^2\overline{S(k)}$ would
approach a constant at small $k$.
Such convergence is clearly seen, for example, in work
on the TISOS model {\em without} disorder \cite{zenghenley}.
For the disordered model, the data does not converge to a
constant over the lengths
we examined; it is instead fit very well by a
logarithmic divergence.
The analysis of the structure factor in this fashion allows us to clearly
distinguish between logarithmically rough and super-rough, since the
difference is singular as $k\rightarrow 0$;
in the $\overline{W^2}(L)$ plots, the
difference between the two predictions is additive.
The logarithmic divergence is precisely what one expects
if $\overline{W^2}(L) \sim \ln^2(L)$.
This is very strong evidence that $B \neq 0$ in Eq.\ (\ref{fitform}).
The prediction of a super-rough ground state is consistent
with our data, while
the prediction of a logarithmically rough phase is inconsistent
with the calculated structure factor.

We also calculated the ground state energy for the fluxlines derived
from the TISOS-model ground states.  We find that the fluctuations in
the energy per flux line, averaged over samples,
$\overline{\Delta E}$
scale as $\overline{\Delta E} \sim L^{0.50\pm0.04}$.

In conclusion, we use a maximum-flow algorithm
and a minimum-cost perfect matching algorithm to find ground state
configurations in the disordered substrate and disordered
TISOS models (which is equivalent
to finding ground states of many flux lines in a random potential).
We find that the ground-state configurations have a $\ln^2(L)$
roughness in both models,
in agreement with the renormalization group calculations
and in disagreement with the predictions of the lowest-order
variational approach.
Our results are much clearer than previous numerical work at finite
temperature \cite{BatrouniHwa,CuleShapir,MarinariEtal},
since the algorithms quickly
find {\em exact} ground states, so that we can average over many
realizations of larger systems and we also study the models far
from $T_g$, so that the coefficient of the $\ln^2(L)$ is larger.
Although the results found here strictly apply to $T=0$ their
agreement with these obtained by RG just below $T_g$ strongly
suggests that the $\ln^2(L)$ behavior holds in the whole $T \le T_g$
glassy phase.
These algorithms can be directly
extended to study flux lines in many different types of disorder,
including columnar and splay defects.

We thank M. Kardar and C. Henley for useful discussions.
C. Zeng received support from NSF grants NMR-9217284 and
NMR-9419257.
A. A. Middleton gratefully acknowledges the Alfred P.\ Sloan
Foundation for support.

\begin{figure}
\caption{Various representations of the TISOS model. The height variables 
defined on each vertex of the triangular lattice (dashed lines) are shown 
in the figure. The equivalent dimer covering is indicated by the thick 
bonds on the dual hexagonal lattice. The corresponding two fluxlines 
are displayed as contiguous triple lines. The linear size of the system 
is denoted by L (=3).
Periodic boundary conditions are imposed (twisted from top to bottom). 
}
\label{fig1}
\end{figure}

\begin{figure}
\caption{
Sample-averaged interface widths for the DS and TISOS models, as
a function of system size $L$, on a logarithmic scale.  The DS
square widths $\overline{W^2}$ have been multiplied by 
a scale factor of 10 for comparison
with the TISOS results.
The fractional
statistical errors in $\overline{W^2}$are less than $0.5\%$;
this error is about half of the size of the symbols in the plot.
The lines are quadratic fits to all of the data shown; residual
errors are within twice the statistical uncertainty for all points.
}
\label{fig2}
\end{figure}

\begin{figure}
\caption{
Sample averaged structure factors for the DS and TISOS models.
The symbols indicate which model and the sample sizes.
$\overline{S(k)}$ has been multiplied by $k^2$ for clearer comparison with the
theoretical predictions, and the curves are normalized so that
$k^2\overline{S(k)} = 1.0$ at $|k|=1$.
The error bars shown
are greater than the statistical error and reflect
the possible error introduced by the smoothing of
the $k^2\overline{S(k)}$ data over a factor of $2$ in $k^2$.
As $k^2\rightarrow 0$, $k^2\overline{S(k)}$ for a
logarithmically rough interface
would approach a constant; the apparent logarithmic divergence of
$k^2\overline{S(k)}$ indicates that the interface is super-rough for the
ground states of these two models.
Note that the finite-size effects are distinct for the two models:
in the DS model, $k^2\overline{S(k)}$ rises above a logarithmic
form for $k$ near
$2\pi/L$ while in the TISOS model, $k^2\overline{S(k)}$ drops below
the logarithmic
form for $k$ near the finite-volume limit.
}
\label{fig3}
\end{figure}

\end{document}